\documentclass[twocolumn,aps,prb]{revtex4} %superscriptaddress,

\usepackage{amsmath,amssymb,amsfonts,bm}
\usepackage{graphicx,epstopdf}  %epsfig,
\usepackage{color}
\usepackage{lscape}
\usepackage{extarrows}
\usepackage{enumitem}
\usepackage{empheq}

 \usepackage{array}
 \usepackage{graphicx}

\allowdisplaybreaks[1]

\renewcommand{\d}{{\rm d}}

\newcommand{\w}{\omega}
\newcommand{\wti}{\widetilde}
\newcommand{\ti}{\tilde}
\newcommand{\B}{\mbox{\tiny B}}

\newcommand{\tT}{\mbox{\tiny T}}

\newcommand{\tS}{\mbox{\tiny S}}

\newcommand{\T}{\mbox{\tiny T}}

\newcommand{\SB}{\mbox{\tiny SB}}

\newcommand{\dg}{\dagger}
\newcommand{\la}{\langle}
\newcommand{\ra}{\rangle}

\newcommand{\Sec}[1]{Sec.\,\ref{#1}}

\newcommand{\nl}{\nonumber \\}
\newcommand{\be}{\begin{equation}}
\newcommand{\ee}{\end{equation}}
\newcommand{\bsube}{\begin{subequations}}
\newcommand{\esube}{\end{subequations}}
\newcommand{\Eq}[1]{Eq.\,(\ref{#1})}
\newcommand{\Eqs}[1]{Eqs.\,(\ref{#1})}
\newcommand{\Fig}[1]{Fig.\,\ref{#1}}

\usepackage{tikz}
\usetikzlibrary{positioning, shapes.geometric}
\newcommand{\RN}[1]{%
  \textup{\rm \uppercase\expandafter{\romannumeral#1}}%
}

\makeatletter
\providecommand{\leftsquigarrow}{%
  \mathrel{\mathpalette\reflect@squig\relax}%
}
\newcommand{\reflect@squig}[2]{%
  \reflectbox{$\m@th#1\rightsquigarrow$}%
}
\makeatother

\begin{document}

\author{Jie Fang}
\thanks{Authors of equal contributions}
\author{Zi-Hao Chen}
\thanks{Authors of equal contributions}

\author{Yu Su}
\author{Zi-Fan Zhu}

\author{Yao Wang}
\email{wy2010@ustc.edu.cn}
\author{Rui-Xue Xu}
\email{rxxu@ustc.edu.cn}

\author{YiJing Yan}

\affiliation{Department of Chemical Physics, University of Science and Technology of China, Hefei, Anhui 230026, China}

\title{
Coherent excitation energy transfer in model photosynthetic reaction center:\\
Effects of non-Markovian quantum environment
}

%\keywords{excitation energy transfer, photosynthetic heat engine, non-Markovian quantum dissipation, quantum coherence}

\begin{abstract}
Excitation energy transfer (EET) and electron transfer (ET) are crucially involved in photosynthetic processes.
In reality, the photosynthetic reaction center constitutes an open quantum system of EET and ET,
which manifests an interplay of pigments, solar light and phonon baths.
So far theoretical studies have been mainly based on master equation approaches in the Markovian condition.
The non-Markovian environmental effect, which may play a crucial role, has not been sufficiently considered.
In this work, we propose a mixed dynamic approach to investigate this open system. The influence of phonon bath is treated
via the exact dissipaton equation of motion (DEOM) while that of photon bath is via the Lindblad master equation.
Specifically, we explore the effect of non-Markovian quantum phonon bath on the coherent transfer dynamics and
its manipulation on the current--voltage behavior.
Distinguished from the results of completely Markovian Lindblad equation and those adopting classical environment description,
the mixed DEOM--Lindblad
 simulations exhibit
transfer coherence up to a few hundreds femtoseconds
and the related environmental manipulation effect on current.
These non-Markovian quantum coherent effects may be extended to
more complex and realistic systems and be helpful to the
design of organic photovoltaic devices.
\end{abstract}

\date{\today}

\maketitle
\newpage

\section{Introduction}
Photosynthesis is one of the most important processes in biological systems, by which plants and other organisms  convert sunlight energy into chemical energy.
It is found that excitation energy transfer (EET) and electron transfer (ET) are crucially involved in the photosynthetic process.
To be concrete, sunlight is absorbed to create an excited state, followed by EET along pigments to reaction center,
where ET happens resulting in charge separation converting excitation energy to chemical energy.

In recent years, the role of quantum coherence in the EET and ET processes
of photosynthesis has got great interest.\cite{Pol220953,Cao204888,Rom17355,Rom14676,Ful14706,Rom104300,Col10644,Eng07782,Lee071462}
The core complex is the main participant in the EET and ET processes of reaction center.
In reality, it constitutes an open quantum system, which manifests an interplay among
pigments, solar light and phonon baths.
The involved dynamics could be non-Markovian in case
that the coupling strength between pigments and the phonon environment be comparable to that
between pigments themselves, as well as the timescale of EET/ET around that of the phonon bath memory.

Unlike the light-harvesting systems, which have been theoretically intensively studied,\cite{Ish09234111,Kre112166,%
Che11194508,Xu13024106,Che1569,Zha16204109,Jan18035003,Kun208783,Yan211498,Kun22349}
dynamics of the EET/ET processes in the reaction center is relatively rarely explored.
So far theoretical studies on EET/ET in the photosynthetic reaction center
are mainly based on some approximate methods in the Markovian condition, such as the Redfield equation,\cite{Rom14676,Wer18084112}
%or its generalized one including both secular and non-secular contributions,\cite{}
polaron master equation,\cite{Qin17012125} Lindblad equation,\cite{Kil15155102} and Pauli master equation.\cite{Cre13253601}
The quantum coherence enhanced effect on electric current was once exhibited
in Ref.\ \onlinecite{Dor132746}.
However, Creatore and co-workers pointed out that
the solutions in Ref.\ \onlinecite{Dor132746}
were unstable and
the numerical evolutions there did not retain the positivity of density matrix,
resulting in artificial behaviors which would diverge with time going on,
see the details in Supplemental Material of Ref.\ \onlinecite{Cre13253601}.
Hence accurate simulations are needed, together with assessments on approximate approaches.

In this work, we study this open system problem using a mixed dynamic approach.
The photon bath (light) influence is treated adopting the
Lindblad equation,\cite{Lin76119,Gor76821}
while that of the phonon environment is via the dissipaton equation of motion (DEOM) method.\cite{Yan14054105,Zha15024112}
The DEOM is a non-Markovian and nonperturbative approach,
constructed on basis of a quasi-particle, dissipaton representation for hybridized collective bath dynamics.
For reduced system dynamics, the DEOM is equivalent to the hierarchical equation of motion (HEOM) formalism,\cite{Tan20020901}
which is established via time derivative on the influence functional path integral
or stochastic fields methods.\cite{Tan89101,Yan04216,Ish053131,Xu05041103,Xu07031107}
Both HEOM and DEOM are exact under Gaussian bath statistics.
The DEOM is more convenient and straightforward to
study environmental dynamics related problems, such as polarizations under external fields.\cite{Zha15024112,Zha16204109,Che21244105}

In numerical demonstrations, the phonon bath will also be treated via the semigroup Lindblad master equation
for comparison with DEOM.
In this way the effects of quantum coherence versus non-Markovian phonon bath will be highlighted.
The remainder of paper is organized as follows.
In \Sec{IIA}, we introduce a five--level model\cite{Rom104300,Dor132746,Kil15155102,Cre13253601,Qin17012125}
applied in our study, which captures the main features of
the EET and ET processes in the photosynthetic reaction center.
The associated Hamiltonian and bath functions are then described.
The mixed DEOM--Lindblad dynamic equations are proposed in \Sec{IIB}.
The construction of Lindblad master equation is briefly outlined in Appendix.
Numerical simulations on transfer dynamics and current--voltage behaviors are demonstrated and discussed in \Sec{num}.
We summarize the paper in \Sec{sum}.

\section{Theoretical description}\label{model}

In this section, we begin with the setup of total system--plus--baths composite exploited in this study, where the system is described by a five--level model.
\cite{Rom104300,Dor132746,Kil15155102,Cre13253601,Qin17012125}
This five--level model system,
based on the photosystem II core complex,
not only characterizes the main features of EET and ET processes, but also takes into account the important interactions involved.
The system Hamiltonian and bath coupling statistics
as well as the proposed mixed DEOM--Lindblad dynamic approach are given after that.
For brevity, we set $\hbar=1$ and $\beta=1/(k_BT)$ throughout the paper, with $k_B$
being the Boltzmann constant and $T$ the temperature.

\begin{figure}
\includegraphics[width=\columnwidth]{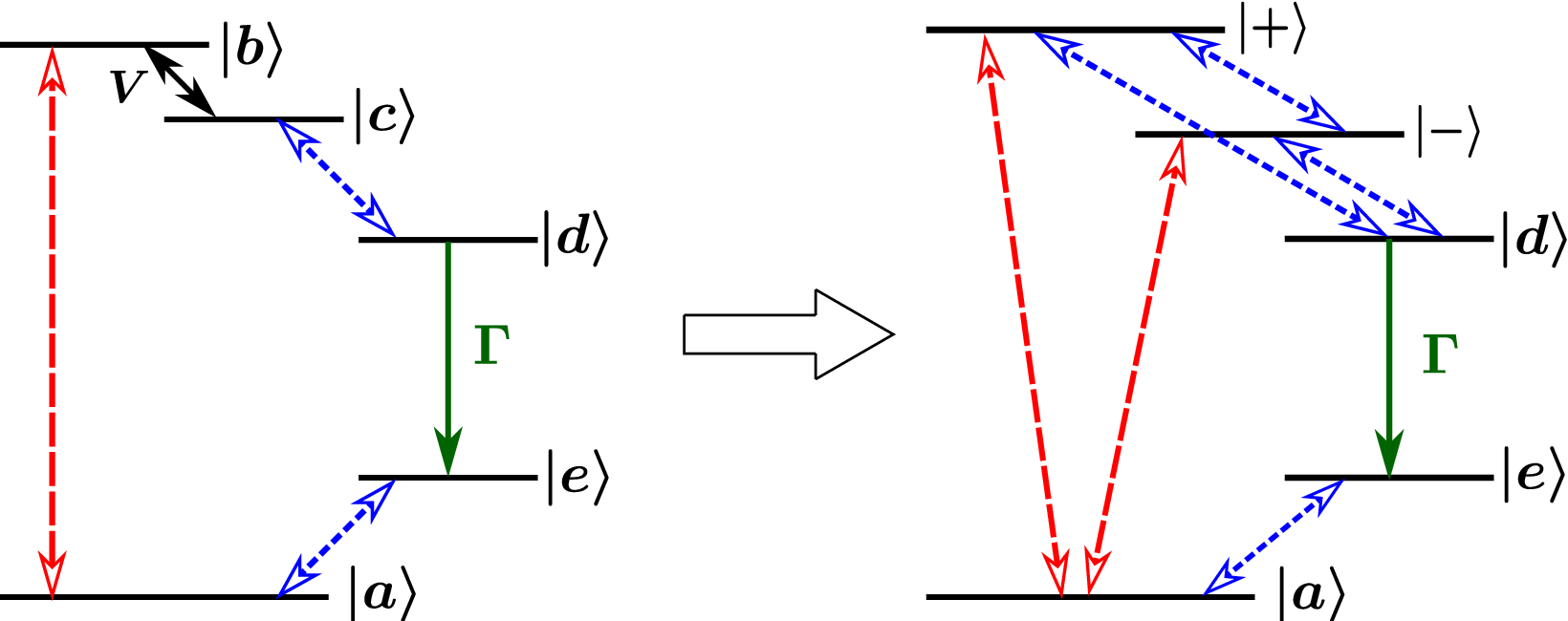}
\caption{Sketch of the total composite model before and after the diagonalization of system Hamiltonian.}
\label{fig1}
\end{figure}

\subsection{Five--site model of system and bath statistics}\label{IIA}
Let us first introduce the
system from the biological perspective.
\cite{Rom104300}
The photosystem II reaction center core complex contains four chlorophylls (special pair P$_{\rm D1}$ and P$_{\rm D2}$ and accessory chlorophylls Chl$_{\rm D1}$ and Chl$_{\rm D2}$) and two pheophytins (Phe$_{\rm D1}$ and Phe$_{\rm D2}$), arranged into two branches (D1 and D2).
Only the D1 branch plays an active role in the photo-induced electron transfer:
\[
  \begin{array}{ccc}
   |a\ra\!: {\rm P}_{\rm D1}{\rm Chl}_{\rm D1}{\rm Phe}_{\rm D1}  & \longrightarrow & ({\rm P}_{\rm D1})^{\ast}{\rm Chl}_{\rm D1}{\rm Phe}_{\rm D1}\!: |b\ra \\
 e \rightsquigarrow\  \uparrow &  &\downarrow \\
  |e \ra\!:{\rm P}_{\rm D1}^{+}{\rm Chl}_{\rm D1}{\rm Phe}_{\rm D1} &  & {\rm P}_{\rm D1}({\rm Chl}_{\rm D1}{\rm Phe}_{\rm D1})^{\ast}\!: |c\ra \\
e \leftsquigarrow \  \uparrow &  &\downarrow \\
      \ \ \ \ \ \ \ {\rm P}_{\rm D1}^{+}{\rm Chl}_{\rm D1}{\rm Phe}_{\rm D1}^{-} & \longleftarrow &\,{\rm P}_{\rm D1}{\rm Chl}_{\rm D1}^{+}{\rm Phe}_{\rm D1}^{-}\ \ \ \ \ \ |d\ra \\
  \end{array}
\]
The D1 branch, ${\rm P}_{\rm D1}{\rm Chl}_{\rm D1}{\rm Phe}_{\rm D1}$, is firstly excited from $|a\ra$ to $|b\ra$ via the absorption of photons. From $|b\ra$ to $|c\ra$  is the EET, followed by the charge separation resulting in state $|d\ra$ where the positive and negative charges are rapidly spatially separated.
For simplicity, a charge--separated state $|d \ra$ is coarsely used to represent both states.
From $|d\ra$ to $|e\ra$, an electron is released from the system.
At last, the system captures an electron from the
surroundings to complete the cycle and returns to the ground
state $|a\ra$.

According to the above description, the system Hamiltonian can be written as
\begin{align}\label{Hsys}
H_{\tS}=\sum_{m\in I}E_{m}|m\ra \la m|+V(|b\ra\la c|+|c\ra\la b|),
\end{align}
with $I\equiv \{a,b,c,d,e\}$.
To phenomenologically describe the electron release
from $|d\ra$ to $|e\ra$,
a superoperator can be introduced as\cite{Dor132746,Kil15155102,Cre13253601,Qin17012125}
\be\label{Lgamma}
L_{\Gamma}{\hat O}=-\frac{\Gamma}{2}\big [{\hat O}|d\ra\la d |+ |d\ra\la d |{\hat O} - 2|e\ra\la d|{\hat O}|d\ra\la e|\big].
\ee
Here, $\Gamma$ is the rate of release.
In square brackets of \Eq{Lgamma}, the last term and the first two terms correspond to
$T_1$ relaxation and $T_2$ dephasing, respectively.\cite{Red651,Yan002068}
%{\color{magenta}
In this work, we exploit this constant rate description as in literature,\cite{Dor132746,Kil15155102,Cre13253601,Qin17012125} to
phenomenologically represent the process from $|d\ra$ to $|e\ra$.
This description is widely applied in various fields, such as chemical kinetics and radioactive decay processes,
where the inverse processes rarely happen. In the photosynthetic reaction center, a series of chemical reactions
are driven by the electron released from $|d\ra$ and finally reach $|e\ra$, while the inverse process from $|e\ra$ to $|d\ra$ is almost prohibited.
%}

For the system--plus--baths composite, the total Hamiltonian reads
\be \label{Ht}
H_{\T}=H_{\tS}+H_{\SB}^{(\rm \uppercase\expandafter{\romannumeral1})}
+H_{\SB}^{(\rm \uppercase\expandafter{\romannumeral2})}
 +h_{\B}^{(\rm \uppercase\expandafter{\romannumeral1})}
 +h_{\B}^{(\rm \uppercase\expandafter{\romannumeral2})}.
\ee
Here, the system Hamiltonian is as in \Eq{Hsys},
while the bath Hamiltonians are
\be
h_{\B}^{(\rm \uppercase\expandafter{\romannumeral1})}
=\sum_{k}\varepsilon_{k}b_{k}^{\dg}b_{k}
\quad {\rm and} \quad
h_{\B}^{(\rm \uppercase\expandafter{\romannumeral2})}
= \sum_{j}        \w_{j}a_{j}^{\dg}a_{j}
\ee
for the photon bath and phonon bath, respectively.  %%, both assumed to be harmonic.
The system-bath interaction Hamiltonians read
\bsube\label{g2}
\begin{align}
H_{\SB}^{(\rm \uppercase\expandafter{\romannumeral1})}
&= \hat Q_1^{(\RN{1})}
   \hat F_1^{(\RN{1})},
\\
H_{\SB}^{(\rm \uppercase\expandafter{\romannumeral2})}
&=\sum_{\mu=2}^{5} \hat Q_{\mu}^{(\RN{2})}
                   \hat F_{\mu}^{(\RN{2})},
\end{align}
\esube
with $\hat Q_1^{(\RN{1})}=|a\ra\la b|+|b\ra\la a|$, $\hat Q_2^{(\RN{2})}=|c\ra\la d|+|d\ra\la c|$,
$\hat Q_3^{(\RN{2})}=|e\ra\la a|+|a\ra\la e|$,
$\hat Q_{4}^{(\RN{2})}=|b\ra\la b|$, and $\hat Q_{5}^{(\RN{2})}=|c\ra\la c|$,
whereas
      $\hat F_{1}^{(\RN{1})}=\frac{1}{\sqrt{2}} \sum_k \ti c_{k}(b_{k}+b_{k}^{\dg})$
and
$
\hat F^{(\RN{2})}_{\mu=2\sim5}=\frac{1}{\sqrt{2}} \sum_j  c_{\mu j}  (a_{j}+a_{j}^{\dg})$.
%%%
These settings are depicted in the left panel of \Fig{fig1} and constitute Gaussian bath couplings.
Their influences on the system can be completely characterized by the spectral densities,
\bsube\be\label{dd3}
  J_{1}^{(\RN{1})}(\w>0)=\frac{\pi}{2}\sum_{k}\ti c_{k}^2  \delta(\w-\varepsilon_{k}),
\ee
and (for $\mu,\nu=2\sim5$)
\be\label{dd2}
J_{\mu\nu}^{(\RN{2})}(\w>0)=\frac{\pi}{2}\sum_{j}c_{\mu j}c_{\nu j}\delta(\w-\w_j).
\ee
\esube

In \Fig{fig1},
red and blue dash arrows represent the state transfers induced by photon and phonon baths, respectively.
The system Hamiltonian eigenstates are $|a\ra$, $|d\ra$, $|e\ra$ and
\begin{align}\label{hello}
 \begin{bmatrix}
  |+\ra \\
  |-\ra
 \end{bmatrix}
 ={\bf U}\begin{bmatrix}
  |b\ra \\
  |c\ra
 \end{bmatrix}
 \equiv
  \begin{bmatrix}
  u_{11}&u_{12} \\
   u_{21}&u_{22}
 \end{bmatrix}
 \begin{bmatrix}
  |b\ra \\
  |c\ra
 \end{bmatrix}
\end{align}
with $\bf U$ being the real and orthogonal transformation matrix which diagonalizes $H_{\tS}$.
Inversely
\begin{align}
 \begin{bmatrix}
  |b\ra \\
  |c\ra
 \end{bmatrix}
 =  \begin{bmatrix}
  u_{11}&u_{21} \\
   u_{12}&u_{22}
 \end{bmatrix}
 \begin{bmatrix}
  |+\ra \\
  |-\ra
 \end{bmatrix}.
\end{align}
Correspondingly, we can recast
\begin{align*}
&\hat Q_1^{(\RN{1})}=u_{11}\big(|a\ra\la +|+|+\ra\la a|\big)+u_{21}\big(|a\ra\la -|+|-\ra\la a|\big),
\\
&\hat Q_2^{(\RN{2})}=u_{12}\big(|+\ra\la d|+|d\ra\la +|\big)+u_{22}\big(|-\ra\la d|+|d\ra\la -|\big),
\\
&\hat Q_{4}^{(\RN{2})}=u^2_{11}|+\ra\la +|+u^2_{21}|-\ra\la -|+u_{11}u_{21}\big(|+\ra\la -|+|-\ra\la +|\big),
\\
&\hat Q_{5}^{(\RN{2})}=u^2_{12}|+\ra\la +|+u^2_{22}|-\ra\la -|+u_{12}u_{22}\big(|+\ra\la -|+|-\ra\la +|\big),
\end{align*}
and $\hat Q_{3}^{(\RN{2})}$ is not affected.
The transformed interaction patterns are exhibited in the right panel of \Fig{fig1}.

\subsection{Mixed DEOM--Lindblad dynamic approach}\label{IIB}
In the total composite space, the total density operator $\rho_{\T}(t)$ evolves as
\be \label{rhoTeom}
\dot\rho_{\T}(t)=-i[H_{\T},\rho_{\T}(t)]+L_{\Gamma}\rho_{\T}(t),
\ee
with $H_{\tT}$ and $L_{\Gamma}$ defined in \Eqs{Ht} and (\ref{Lgamma}), respectively.
In the proposed mixed dynamic approach,
light is treated as photon bath via the Lindblad master equation, detailed in Appendix.
Thus an additional superoperator for the action of light is now introduced as
\be\label{eq10}
L^{(\RN{1})}=\gamma_{+}u_{11}^2L_{+}+\gamma_{-}u_{21}^2L_{-}\,,
\ee
with $\gamma_\pm\equiv2J_{1}^{(\RN{1})}(\w_{\pm a})$ the dissipative rate and
\begin{align}\label{eq11}
L_{\pm}\hat O&=(1+\bar n_{\pm})\Big(
  \la \pm |\hat O|\pm \ra\,|a   \ra\la a|   -\frac{1}{2}\big\{|\pm\ra\la \pm|,\hat O\big\}\Big)
\nl &\quad
+\bar n_{\pm} \Big(
  \la a   |\hat O|a   \ra\,|\pm \ra\la \pm |-\frac{1}{2}\big\{|a  \ra\la a  |,\hat O\big\}\Big),
\end{align}
where $\bar n_\pm\equiv\bar n_{\pm a}$; cf.\ Appendix.

To explore non-Markovian and non-perturbative
influence of phonon bath, we adopt the well-established DEOM approach.
It starts with the exponential expansion form of bath coupling correlation functions,
\be\label{FDT}
\wti C_{\mu\nu}^{(\RN{2})}(t)
=\frac{1}{\pi}\int^{\infty}_{-\infty}
  \d\w \frac{e^{-i\w t}  J_{\mu\nu}^{(\RN{2})}(\w)}{1-e^{-\beta\w}}
=\sum_\kappa\xi^{\mu\nu}_{\kappa}e^{-\gamma^{\mu\nu}_{\kappa}t}\,.
\ee
The first identity is the fluctuation--dissipation theorem.\cite{Wei21,Yan05187}
The standard DEOM algebra gives rise to\cite{Yan14054105,Zha15024112}
\begin{align}\label{DEOM}
 \dot\rho^{(n)}_{\bf n}&=
 -\Big[i{\cal L}_{\tS}-L_{\Gamma}-L^{(\RN{1})}+\sum_{\mu\nu\kappa} n^{\mu\nu}_{\kappa}\gamma^{\mu\nu}_{\kappa}\Big]\rho^{(n)}_{\bf n}
\nl&\quad
  -i\sum_{\mu\nu \kappa}\Big[{\cal A}_{\mu}\rho^{(n+1)}_{{\bf n}_{\mu\nu \kappa}^+}
% \nl&\quad
   +n^{\mu\nu}_{\kappa}{\cal C}^{\mu\nu}_{\kappa}
   \rho^{(n-1)}_{{\bf n}_{\mu\nu \kappa}^-}\Big],
\end{align}
with ${\cal L}_{\tS}\hat O\equiv[H_{\tS},\hat O]$ and
\bsube
\begin{align}
{\cal A}_{\mu}\hat O&\equiv[\hat Q_{\mu}^{(\RN{2})},\hat O],\\
{\cal C}^{\mu\nu}_{\kappa}\hat O&\equiv\xi^{\mu\nu}_{\kappa}\hat Q_{\nu}^{(\RN{2})}\hat O
-\big(\xi^{\mu\nu}_{\bar\kappa}\big)^{\ast}\hat O\hat Q_{\nu}^{(\RN{2})}.
\end{align}
\esube
This is the mixed DEOM--Lindblad formalism.
The term of index $\bar\kappa$ is associated with that of $\kappa$
by $\gamma^{\mu\nu}_{\bar\kappa}\equiv(\gamma^{\mu\nu}_{\kappa})^\ast$.
The indices of density matrices are denoted as $\textbf{n}=\{n^{\mu\nu}_{\kappa}\}$,
an ordered set of the bosonic dissipaton's occupation numbers, $n^{\mu\nu}_{\kappa}=0,1,\cdots$,
and $n=\sum_{\mu\nu\kappa}n^{\mu\nu}_{\kappa}$ the total number.
 ${\bf n}^{\pm}_{\mu\nu\kappa}$ differs from ${\bf n}$ only
at the specified
$n^{\mu\nu}_{\kappa}$ by $\pm 1$.
$\rho^{(0)}_{\bf 0}$ is just the reduced system density operator, while the others, $\rho^{(n\geq 1)}_{\bf n}$,
coupled to $\rho^{(0)}_{\bf 0}$ in a hierarchical manner, are
dissipaton density operators.

\section{Numerical demonstrations and discussions}\label{num}
	For numerical simulations, we adopt the Drude model for the phonon bath spectral densities (for $\mu,\nu=2\sim5$),
\be\label{drude}
  J_{\mu\nu}^{(\RN{2})}(\w)=\frac{2\eta_{\mu\nu}\lambda_{\nu}\gamma_{\nu}\w}{\w^2+\gamma_{\nu}^2}\,.
\ee
The Drude model is a strongly overdamped solvent model.
In \Eq{drude}, $\lambda_{\nu}$ is the reorganization energy and
$\gamma_{\nu}$ is the damping rate.
$\{\eta_{\mu\nu}\}$ should form a positive--definite matrix.
It characterizes the correlation between different dissipative modes, cf.\ \Eq{dd2}.
Here, the way of denoting cross--correlation contributions is
essentially the same as in Ref.\ \onlinecite{Ish10055004}.
We set parameters as in Table.\,\ref{para}.
They are selected in accordance with Refs.\ \onlinecite{Kil15155102,Cre13253601,Dor132746}.
Particularly, the photon average occupation paramter, $\bar n_{\pm}$, is chosen to be $60000$, to match with Refs.\ \onlinecite{Kil15155102,Cre13253601,Dor132746}.
As pointed out in Ref.\ \onlinecite{Kil15155102}, this represents solar energy concentration within the antenna and
is not related to the actual physical temperature of photon bath.
%excited from $|a\ra$ to $|b\ra$ via the absorption of photons.
%
\begin{table}[!ht]
    \centering
    \begin{tabular}{|m{2cm}<{\centering}|m{2cm}<{\centering}|m{2cm}<{\centering}|}
    \hline
        Parameters & Units & Values \\ \hline
        $E_a$ & cm$^{-1}$ & 0 \\ %\hline
        $E_b$ & cm$^{-1}$ & 14856 \\ %\hline
        $E_c$ & cm$^{-1}$ & 14736 \\ %\hline
        $E_{d}$ & cm$^{-1}$ & 13245 \\ %\hline
        $E_{e}$ & cm$^{-1}$ & 1611 \\ %\hline
        $V$ & cm$^{-1}$ & 30 \\ %\hline
        $\gamma_{\pm}$ & cm$^{-1}$ & 0.005 \\ %\hline
        $\bar n_{\pm}$ & ~ & 60000\\ %\hline
 %               $\bar n_{-,a}$ & ~ & 60000\\ %\hline
        $T$ & K & 300 \\ %\hline\\
        $\lambda_2$ & cm$^{-1}$ & 140 \\ %\hline
        $\gamma_2$ & cm$^{-1}$ & 140 \\ %\hline
        $\lambda_3$ & cm$^{-1}$ & 200 \\ %\hline
        $\gamma_3$ & cm$^{-1}$ & 200 \\ %\hline\\
        $\lambda_4$ & cm$^{-1}$ & 100 \\ %\hline\\
         $\gamma_4$ & cm$^{-1}$ & 10 \\ %\hline\\
        $\lambda_5$ & cm$^{-1}$ & 100 \\ %\hline\\
         $\gamma_5$ & cm$^{-1}$ & 10 \\ %\hline\\
\hline
    \end{tabular}
    \caption{Parameters used in the simulations.}\label{para}
\end{table}
The setup of system Hamiltonian results in the following ${\bf U}$--matrix in \Eq{hello},
\be\nonumber
{\bf U}=\begin{bmatrix}
  0.973 & -0.230 \\
 0.230 & 0.973
 \end{bmatrix}.
\ee
In simulations, the $\{\eta_{\mu\nu}\}$ parameters are set as
\be\nonumber
 \begin{bmatrix}
  \eta_{22} & \eta_{23} & \eta_{24} & \eta_{25}  \\
  \eta_{32} & \eta_{33} & \eta_{34} & \eta_{35}  \\
  \eta_{42} & \eta_{43} & \eta_{44} & \eta_{45}  \\
  \eta_{52} & \eta_{53} & \eta_{54} & \eta_{55}
 \end{bmatrix}
=
 \begin{bmatrix}
  1 & 0 & 0 & 0 \\
  0 & 1 & 0 & 0 \\
  0 & 0 & 1 & \eta  \\
  0 & 0 & \eta & 1
 \end{bmatrix}
.
\ee
We choose $\eta=1$ and $\eta=-1$ to represent the fully correlated and anti--correlated scenarios of
the involved two fluctuating modes, $\hat Q_{4}^{(\RN{2})}=|b\ra\la b|$ and $\hat Q_{5}^{(\RN{2})}=|c\ra\la c|$,
respectively.
Note their real effects shall be considered with the eigenstates of system
and may be different if the sign and value of the coherent coupling $V$ change.
The $\Gamma$ parameter in \Eq{Lgamma} will be varied in the following demonstration.

%{\color{blue}
Before numerical discussions, it is worth to clarify that
non-Markovian quantum nature
is basically the real physical characteristics of phonon environments.
Whether the caused effect is important or not theoretically depends on parameters.\cite{Xu079618,Din12224103}
When approximate treatments are adopted, validity is to be assessed by comparing with accurate methods.
Meanwhile the existing difference in such comparisons would reflect the feature of the neglected factor in those approximate approaches.
In the following part of this section,
the completely Markovian Lindblad equation or methods adopting classical environment description
will be assessed via comparison to the simulation results given by the mixed DEOM--Lindblad approach.
The effects of non-Markovian quantum phonon environment can then be analyzed in due course.
%}

\begin{figure}
\includegraphics[width=\columnwidth]{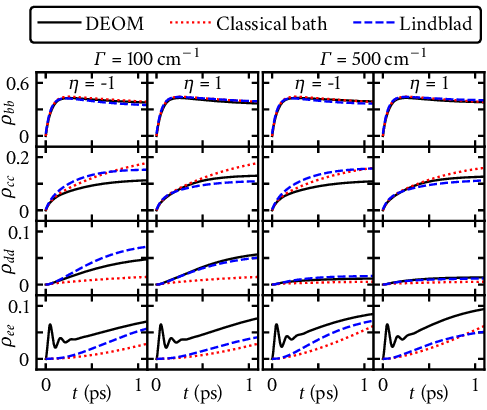}
\caption{Population evolutions evaluated via the mixed DEOM--Lindblad (black-solid),
DEOM--Lindblad under the classical bath limit (red-dot), and complete Lindblad (blue-dash) methods
with varied $\Gamma$ and $\eta$ parameters.}
\label{fig2}
\end{figure}

\begin{figure}
\includegraphics[width=\columnwidth]{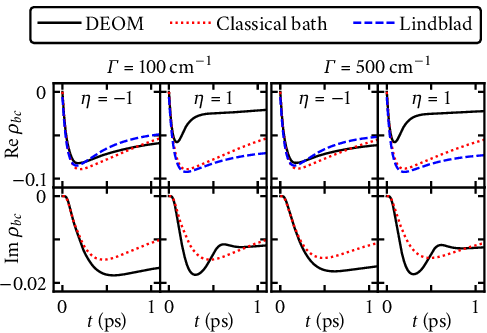}
\caption{Decoherence during EET between the states $|b\ra$ and $|c\ra$,
evaluated via the mixed DEOM--Lindblad (black-solid), DEOM--Lindblad under the classical bath limit (red-dot),
and complete Lindblad (blue-dash) methods
with varied $\Gamma$ and $\eta$ parameters.}
\label{fig2add}
\end{figure}

Figures \ref{fig2} and \ref{fig2add} depict the transient dynamics obtained from different approaches.
We choose cases where $\Gamma=100$ and $500$\,cm$^{-1}$,
both simulated for $\eta=-1$ and $\eta=1$. The system is set to be at $|a\ra$ initially.
The black-solid, DEOM curves are evaluated via the mixed DEOM--Lindblad formalism proposed in \Sec{IIB}.
Both the expansion of phonon bath correlation functions and the hierarchy of dynamic equations are converged.
To explore the quantum phonon environment effects, results from the classical bath correspondence
are illustrated with the red-dot curves for comparison.
In the classical bath condition, the involved bath correlations are real functions.
Thus, the difference of red-dot curves to the black-solid ones
is due to neglecting the imaginary parts of the second identity of \Eq{FDT}.
The blue-dash, Lindblad results are obtained by applying the Lindblad master equation
for both photon and phonon bath operations, cf.\ \Eqs{A10}--(\ref{A11}) of Appendix.
Note the Pauli master equation adopted in Ref.\ \onlinecite{Cre13253601}
neglects the off-diagonal elements of system density matrix. It corresponds to
the Lindblad equation of \Eq{A10}
subject to a population projection.
Therefore the non-Markovian correlated environmental effects can be highlighted in comparison between black-solid and blue-dash curves.

\begin{figure}
\includegraphics[width=\columnwidth]{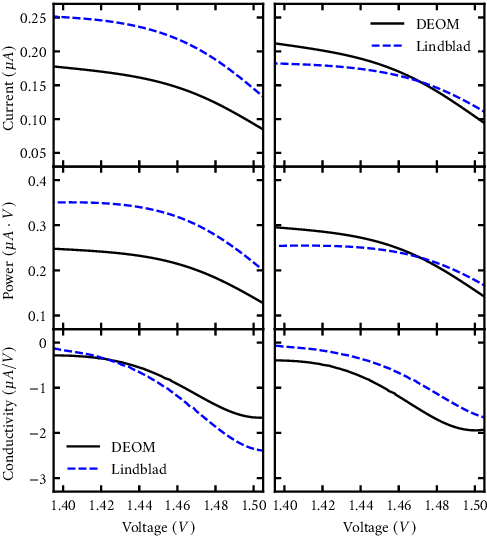}
\caption{Current, power, and conductivity versus the effective voltage with the $\Gamma$ parameter varied.
The black solid curves are from the mixed DEOM--Lindblad simulations while the blue dash curves are from the complete Lindblad simulations.
}
\label{fig3}
\end{figure}

In \Fig{fig2}, the mixed DEOM--Lindblad (black-solid) simulations exhibit
quantum coherence in the panels of $\rho_{ee}$.
Both the complete Lindblad (blue-dash) and classical bath (red-dot) results show little quantum coherent behavior,
similar as the time evolutions demonstrated in Supplemental Material of Ref.\ \onlinecite{Cre13253601}.
The associated decoherence processes of $\rho_{bc}$ are depicted in \Fig{fig2add}.
There are no Lindblad (blue-dash) curves in the panels of ${\rm Im}\rho_{bc}$,
because its results retain zero from the chosen initial state.
It is observed that 
by the mixed DEOM--Lindblad (black-solid) simulations,
the anti--correlated bath fluctuations with $\eta=-1$ lead to faster dephasing processes than
the correlated ones with $\eta=1$, for the present system.  %as observed from \Fig{fig2add}.
In both \Fig{fig2} and \Fig{fig2add}, the $\Gamma=100$ and $500$\,cm$^{-1}$ cases give similar transient behaviors except for $\rho_{dd}$ [cf.\ \Eq{Lgamma} and comments after it].
Further effects of $\Gamma$ and $\eta$ on steady states and the associated current--voltage properties are to be demonstrated in \Fig{fig3}.

As in the literature,\cite{Rom104300,Dor132746,Kil15155102,Cre13253601,Qin17012125}
 we may view the composite as a biological heat engine, with the steady-state current,
\be\label{current}
j=e\Gamma \rho_{\tS;dd}^{\rm st}\,,
\ee
and the effective voltage $\Phi$ via
\be\label{voltage}
e\Phi=E_{d}-E_{e}+k_{B}T\ln\frac{\rho_{\tS;dd}^{\rm st}}{\rho_{\tS;ee}^{\rm st}}\,.
\ee
Here, $e$ is the electron charge.
Figure \ref{fig3} depicts the current (upper-panels), power $j\cdot\Phi$ (middle-panels), and
conductivity $\d j/\d\Phi$ (lower-panels) versus the voltage,
for the results from the mixed DEOM--Lindblad (black-solid) and complete Lindblad (blue-dash) simulations.
The current and voltage, $j$ and $\Phi$, are evaluated from \Eq{current} and \Eq{voltage}, respectively, with the $\Gamma$ parameter
varied from 600 down to 8\,cm$^{-1}$ for the mixed DEOM--Lindblad,
and 900 down to 12\,cm$^{-1}$ for the complete Lindblad simulations.
Note that under the classical bath condition, $\rho_{\tS;dd}^{\rm st}=0$, leading to both current and voltage undefined.
Negative conductivity is observed owing to the setup of the present model heat engine.
%%The Lindblad exhibits some illusive unstable behaviors.

In \Fig{fig3}, the mixed DEOM--Lindblad results show certain manipulation %switch analogue,
effects by adjusting the cross--correlation, $\eta$--parameter, between different environmental couplings.
The current evaluated via the mixed DEOM--Lindblad simulation (black-solid) is overall enlarged in the $\eta=1$ case
(upper--right panel) compared with $\eta=-1$ (upper--left panel).
This observation highlights the relationship among the non-Markovian quantum environment, the transfer coherence, and the current enhancement.
Recall that we have shown in \Fig{fig2add} the former case possesses a longer time of coherence than the latter one.
In contrast, since the non-Markovianity and quantum coherence are not fully covered in the complete Lindblad approach,
 it produces the opposite behaviors
 that the current is weakened in the $\eta=1$ case in comparison with $\eta=-1$, as
seen from the blue-dash curves in the upper panels of \Fig{fig3}.

\section{Summary}\label{sum}

In this work, we propose a mixed DEOM--Lindblad approach to study the transient dynamics and steady-state
current--voltage behaviors of a model photosynthetic reaction center system.
The photon bath (light) influence is treated via the Lindblad dissipative superoperator
while that of the phonon environment is via the exact DEOM method taking into account the
non-Markovian and non-perturbative effects.
The correlation between photon and phonon baths' couplings on the reduced system are also included in
the construction of the mixed DEOM--Lindblad formalism.
The transfer dynamics and steady-state current--voltage behaviors are compared among different approaches,
the mixed DEOM--Lindblad, complete Lindblad, and DEOM--Lindblad with classical bath limit,
to explore the non-Markovian quantum environment effects.
Distinguished from the other two methods, results via the mixed DEOM--Lindblad simulation
exhibit the transfer coherence up to a few hundreds femtoseconds
and an environment manipulation effect on the current enhancement.
As DEOM is an accurate method, the present observations of non-Markovian quantum coherent effects
are expected to be extended to more complex and realistic systems and be helpful to the design of organic photovoltaic devices.

\begin{acknowledgments}
Support from the Ministry of Science and Technology of China (Nos.\ 2017YFA0204904 and 2021YFA1200103),
the National Natural Science Foundation of China (Nos.\ 22103073 and 22173088),
and Anhui Initiative in Quantum Information Technologies
is gratefully acknowledged. Y.\ Wang  and Z.\ H.\ Chen thank also the
partial support from GHfund B (20210702).
\end{acknowledgments}

\appendix
\section*{Appendix: Constructional detail of Lindblad master equation}
\label{appa}

In this appendix, we give the constructional detail
of the Lindblad master equation.
Consider a general form of system--plus--bath total Hamiltonian,
\be \label{Htapp}
H_{\T}=H_{\tS}+\sum_{\mu}\hat Q_{\mu}^{\tS}\hat F^{\B}_{\mu}+h_{\B}.
\ee
The time-local quantum dissipation equation for the reduced system density operator,
via the cumulant partial ordering prescription with neglecting bath dispersion,
is obtained as\cite{Yan982721,Yan002068}
\be \label{red}
 \dot \rho_{\tS}(t)=-i{\cal L}_{\tS}\rho_{\tS}(t)-\sum_{\mu}[\hat Q^{\tS}_\mu,
  \wti Q_{\mu}\rho_{\tS}(t)-\rho_{\tS}(t)\wti Q_{\mu}^{\dg}],
\ee
with %%${\cal L}_{\tS}\hat O=[H_{\tS},\hat O]$,
\be
\wti Q_{\mu}\equiv\sum_{\nu}C_{\mu\nu}(-{\cal L}_{\tS})\hat Q^{\tS}_{\nu},
\ee
and
\be
C_{\mu\nu}(\w)\equiv\frac{1}{2}\int^{\infty}_{-\infty}{\rm d}\tau\, e^{i\w \tau}\wti C_{\mu\nu}(\tau)
 =[C_{\nu\mu}(\w)]^{\ast}.
\ee

To obtain the concrete form of Lindblad equation,
we shall recast $\hat Q^{\tS}_\mu$ and $\wti Q_{\mu}$ in the system eigenstate representation,
$\{|m\ra\}$ satisfying $H_{\tS}|m\ra=\epsilon_m |m\ra$, as
\bsube
\begin{align} \label{lin0}
&\hat Q^{\tS}_\mu=\sum_{mn}Q^{\tS}_{\mu;mn}|m\ra\la n|,
\\  %\quad {\rm and}\quad
&\wti Q_{\mu}=\sum_{\nu mn}C_{\mu\nu}(\w_{nm})Q^{\tS}_{\nu;mn}|m\ra\la n|,
\end{align}
\esube
with
\be
\w_{mn}\equiv \epsilon_m-\epsilon_n
\ \ {\rm and}\ \
Q^{\tS}_{\mu;mn}\equiv \la m|\hat Q^{\tS}_\mu|n\ra.
\ee
We obtain
\begin{align}\label{lin1}
\dot \rho_{\tS}(t)=-i{\cal L}_{\tS}\rho_{\tS}(t)+
\!\! \sum_{\substack{\mu\nu mnm'n'}} \!\!
\left[(\RN{1})-(\RN{2})-(\RN{3})\right],
 \end{align}
with
%\bsube\label{app123}
\begin{align}
&(\RN{1})=[C_{\mu\nu}(\w_{mn})+C_{\mu\nu}(\w_{n'm'})]\hat S_{\nu;m'n'}\rho_{\tS}(t)\hat S_{\mu;nm}^{\dg},
\nl
&(\RN{2})=C_{\mu\nu}(\w_{n'm'})\hat S_{\mu;nm}^{\dg}\hat S_{\nu;m'n'}\rho_{\tS}(t),
\nl
&(\RN{3})=C_{\mu\nu}(\w_{mn})\rho_{\tS}(t)\hat S_{\mu;nm}^{\dg}\hat S_{\nu;m'n'}.   %\nonumber
\end{align}
%\esube
Here,
$\hat S_{\mu; mn}\equiv Q^{\tS}_{\mu; mn}|m\ra\la n|$,
satisfying
\be
 \hat S_{\mu; mn}^{\dg}=Q^{\tS}_{\mu; nm}|n\ra\la m|=\hat S_{\mu; nm}.
\ee

Now applying the rotating wave approximation
that only terms of $n'=m$ and $m'=n$ contribute, \Eq{lin1} gives rise to
\begin{align}
\dot \rho_{\tS}(t)&=-i{\cal L}_{\tS}\rho_{\tS}(t)+\sum_{\substack{\mu\nu mn}}C_{\mu\nu}(\w_{mn})
\Big[2 \hat S_{\nu;nm}\rho_{\tS}(t)\hat S_{\mu;nm}^{\dg}
\nl &
\quad -\hat S_{\mu;nm}^{\dg}\hat S_{\nu;nm}\rho_{\tS}(t)
-\rho_{\tS}(t)\hat S_{\mu;nm}^{\dg}\hat S_{\nu;nm}\Big].
\end{align}
The detailed-balance relation reads
\be
 C_{\mu\nu}(\w)=J_{\mu\nu}(\w)[1+\bar n(\w)]=J_{\nu\mu}(-\w)\bar n(-\w).
\ee
Note that $J_{\nu\mu}(-\w)=-J_{\mu\nu}(\w)$
and $\bar n(\w)+\bar n(-\w)=-1$
where $\bar n(\w)=1/(e^{\beta\w}-1)$.
We obtain readily
\begin{align}\label{A10}
\dot \rho_{\tS}(t)&
=\Big[\!-i{\cal L}_{\tS}+\sum_{\mu\nu mn}\big(L_{\mu\nu mn}^{(+)}+L_{\mu\nu mn}^{(-)}\big)\Big]\rho_{\tS}(t)
\end{align}
where [$\bar n_{mn}\equiv \bar n(\w_{mn})$]
\bsube\label{A11}
\begin{align}
L_{\mu\nu mn}^{(+)}\hat O&=\frac{1}{2}J_{\mu\nu}(\w_{mn})(1+\bar n_{mn})
\Big(2 \hat S_{\nu;nm}\hat O\hat S_{\mu;nm}^{\dg}
\nl & \quad
-\hat S_{\mu;nm}^{\dg}\hat S_{\nu;nm}\hat O
-\hat O\hat S_{\mu;nm}^{\dg}\hat S_{\nu;nm}\Big),
\\
L_{\mu\nu mn}^{(-)}\hat O&=\frac{1}{2}J_{\nu\mu}(\w_{mn})\bar n_{mn}\Big(
2 \hat S_{\nu;nm}^{\dg}\hat O\hat S_{\mu;nm}
\nl &\quad
-\hat S_{\mu;nm}\hat S_{\nu;nm}^{\dg}\hat O
-\hat O\hat S_{\mu;nm}\hat S_{\nu;nm}^{\dg}\Big).
\end{align}
\esube
This is just the standard form of Lindblad master equation.\cite{Lin76119,Gor76821}
It is also equivalent to the secular Redfield equation.\cite{Red651,Yan002068}

In comparison with DEOM for the non-Markovian influence of phonon bath,
the Markovian Lindblad master equation treatment in \Sec{num} is as the above \Eq{A10} with \Eq{A11}.
Note that the system eigenstate representation shall be adopted.
For the photon bath with the single coupling mode $\hat Q_1^{(\RN{1})}$ in \Sec{IIA},
we finally obtain \Eq{eq10} with \Eq{eq11}.

%\bibliographystyle{../aiptit}
%\bibliography{../bibrefs}

\end{document}